\documentclass[aps,prl,twocolumn,groupedaddress]{revtex4}
\usepackage{amssymb}
\usepackage{graphicx}

\begin{document}

\title{Doping-insensitive density-of-states suppression in polycrystalline iron-based superconductor SmO$_{1-x}$F$_{x}$FeAs}

\author{H. W. Ou$^{1}$, Y. Zhang$^{1}$, J. F. Zhao$^{1}$, J. Wei$^{1}$,  D. W. Shen$^{1}$, B. Zhou$^{1}$,
L. X. Yang$^{1}$, F. Chen$^{1}$, M. Xu$^{1}$,  C. He$^{1}$, R. H.
Liu$^2$, M.~Arita$^3$, K. Shimada$^3$, H.~Namatame$^3$,
M.~Taniguchi$^3$, Y. Chen$^{1}$, X. H. Chen$^2$, and D. L.
Feng$^{1}$} \email{dlfeng@fudan.edu.cn}

\affiliation{$^1$Department of Physics, Surface Physics Laboratory
(National Key Laboratory), and Advanced Materials Laboratory, Fudan
University, Shanghai 200433, P. R. China}

\affiliation{$^2$Hefei National Laboratory for Physical Sciences at
Microscale and Department of Physics, University of Science and
Technology of China, Hefei, Anhui 230026, P. R. China}

\affiliation{$^3$Hiroshima Synchrotron Radiation Center and Graduate
School of Science, Hiroshima University, Hiroshima 739-8526, Japan}

\date{\today}

\begin{abstract}

We investigated the temperature dependence of the density-of-states
in the iron-based superconductor SmO$_{1-x}$F$_x$FeAs ($x=0, 0.12,
0.15, 0.2$) with high resolution angle-integrated photoemission
spectroscopy. The density-of-states suppression is observed with
decreasing temperature in all samples, revealing two characteristic
energy scales (10\,meV and 80\,meV). However, no obvious doping
dependence is observed. We argue that  the 10\,meV suppression is
due to an anomalously doping-independent normal state pseudogap,
which becomes the superconducting gap once in the superconducting
state; and alert the possibility that the 80\,meV-scale suppression
might be an artifact of the polycrystalline samples.
\end{abstract}

\maketitle

The recent discovery of superconductivity in the iron oxypnictides
LnO$_{1-x}$F$_x$FeAs (Ln=La, Sm, Nd, \textit{etc.}) with  transition
temperatures ($T_{c}$'s) well beyond the McMillan limit for BCS
superconductors   has rejuvenated intensive research on
unconventional superconductivity
\cite{JACS,ChenNature,NLWang1,ZXZhao1,ZXZhao2,McMillan,media}.  In
some ways, the  layered structure of a iron-based superconductor
resembles the high-T$_c$ cuprates: LnO$_{1-x}$F$_x$ layers act as
the charge reservoir, and FeAs layers act as the conducting layers.
Moreover, neutron diffraction measurements show that the ground
state of the parent compound LaOFeAs is a spin density wave (SDW)
\cite{PCDai}. This makes the phase diagrams of cuprates and iron
oxypnictides share some common features as well. Whether or not the
superconductivity in iron oxypnictides follows the same physics as
in cuprates thus becomes a fundamental question to ask.

One prominent feature for cuprates is the pseudogap, \textit{i.e.}
suppression of density-of-states (DOS) in the normal state. For
LaO$_{1-x}$F$_x$FeAs, several nuclear magnetic resonance experiments
have argued the existence of a pseudogap \cite{NMR1,NMR2,NMR3}.
Recently, Sato \emph{et al.} \cite{takahashi} reported the angle
integrated photoemission spectroscopy (AIPES) results on
polycrystalline LaO$_{0.93}$F$_{0.07}$FeAs ($T_c$=24K), and they
observed a pseudogap of $15\sim20$ meV far above $T_c$ up to 130K,
together with a finite DOS at the Fermi energy ($E_F$). Almost the
same time, Ishida \emph{et al.} \cite{shin} reported AIPES evidence
for a 100\,meV pseudogap-like feature at 250K in
LaO$_{0.89}$F$_{0.11}$FeAs (LOFFA, $T_c$=26K), which shrinks to
20\,meV at 70K. Moreover, their data on the low-$T_c$
LaO$_{0.94}$F$_{0.06}$FeP (LOFFP, $T_c$=5K) shows a less pronounced
20\,meV pseudogap-like feature. Therefore, these experiments on
polycrystals do suggest the existence of a pseudogap of $9\sim46 k_B
T_c$, and it is very similar to the $10\sim40 k_B T_c$ pseudogap
scale observed in cuprate superconductor\cite{RMP}. At the lowest
temperature, both groups observed a spectral weight suppression
which was related to superconducting gaps of about $4\sim5 meV$ for
LOFFA's, and a superconducting gap of about 1 \,meV for LOFFP.
Intriguingly, they are all about $2k_B T_c$.

\begin{figure}[b!]
\includegraphics[width=7cm]{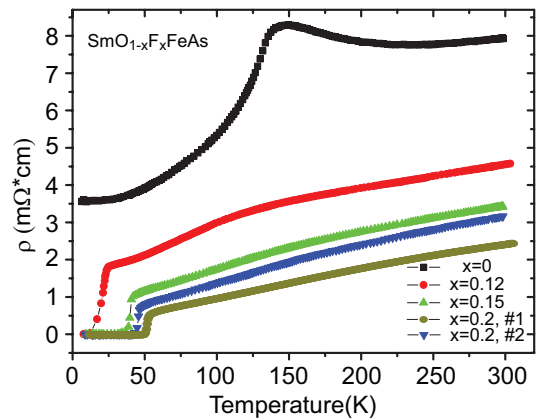}
\caption{(color online). Temperature dependence of Resistivity for
SmO$_{1-x}$F$_{x}$FeAs (x=0, 0.12, 0.15, 0.2)(data are taken from
Ref.[15]).}
\end{figure}

Photoemission is sensitive to surface and sample quality,
systematics thus needs to be collected before reaching a robust
conclusion when dealing with polycrystals. However, so far only one
doping of each material was measured in individual studies. In this
paper, we report systematic AIPES measurement of polycrystalline
SmO$_{1-x}$F$_{x}$FeAs (SOFFA) with a variety of dopings (x=0, 0.12,
0.15, 0.2). At high temperatures, we observed a DOS suppression over
a large energy scale of 80\,meV, whose onset temperature is even
higher than 300K. At low temperatures, a V-shaped lineshape around
the energy scale of 10\,meV was observed in the symmetrized spectra.
Surprisingly, we found that both kinds of suppressions are
doping-independent, very different from the  pseudogap behaviors in
cuprates. The low energy suppression scale of 10\,meV corresponds to
$2k_B T_c^{max}$,  where $T_c^{max}$ is the maximum $T_c$ of 54K for
SOFFA's. Considering similar correlations in  LOFFA and LOFFP, it
suggests that the 10\,meV suppression reflects an intrinsic
pseudogap in the normal state, which evolves into the
superconducting gap at low temperatures. However, it is not clear at
this stage whether the 80meV scale is related to an intrinsic
pseudogap effect, or caused by extrinsic effects such as
inhomogeneities and domain boundaries of the polycrystal.

Polycrystalline SmO$_{1-x}$F$_{x}$FeAs (x=0, 0.12, 0.15, 0.2) have
been synthesized through solid state reaction, the detailed
information about the synthesis and characterization of the sample
has been described elsewhere\cite{ChenXH2}. Resistivity data (Fig.1)
clearly indicate the superconducting transition occurs at 25K, 42K,
47K, and 54K for x=0.12, 0.15, 0.2(\#2), and 0.2(\#1) respectively.
The Meissner volumes were estimated to be $50\sim60$\% according to
the susceptibility measurements, which is among the best for the
polycrystalline samples synthesized so far. The resistivity of the
parent compound  exhibits a similar peak at around 150K similar to
that observed in LOFFA, indicative of a possible SDW and/or
structure phase transition in SOFFA \cite{PCDai}.

\begin{figure}[t]
\includegraphics[width=8cm]{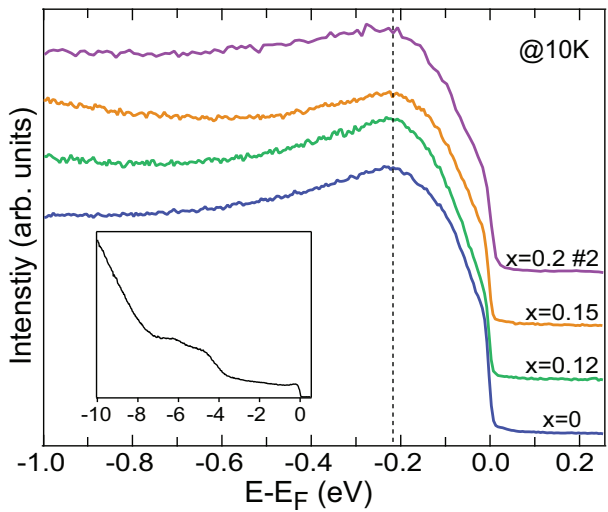}
\caption{(color online).  AIPES spectrum near $E_F$ as a function of
doping for SmO$_{1-x}$F$_{x}$FeAs. The inset shows AIPES spectrum of
SmO$_{0.8}$F$_{0.2}$FeAs over a large energy window. Data were taken
with 22.7eV photons at 10K. }
\end{figure}

Photoemission measurements were performed at beam line 5-4 of
Stanford Synchrotron Radiation Laboratory (SSRL) and beam line 9 of
Hiroshima Synchrotron Radiation Center (HiSOR), data were taken with
Scienta R4000 electron analyzers. The overall energy resolution was
set to 7\,meV. The sample rod was cracked \textit{in-situ }and then
measured in ultra-high vacuum ($\sim 3\times 10^{-11}\,mbar$). We
emphasize that all the sample surfaces prepared in this way show
consistent results, and the data measured with 22.7\,eV photons
agree with those measured with more bulk sensitive 8eV photons
(shown in Fig.4), indicative of the high sample quality.

The AIPES spectra of SmO$_{1-x}$F$_{x}$FeAs are shown in Fig.\,2,
which measure the DOS. There is a broad low energy feature around
0.22\,eV below $E_F$, whose position seems to be quite doping
independent. The low energy feature was attributed to Fe $3d$ states
in various band calculations\cite{LDA0,LDA1,LDA2,LDA3,LDA4}. Our
data, especially the peak at 0.22\,eV and the flat DOS between
0.5\,eV to 3\,eV binding energy (see inset of Fig.2)\cite{Ou1},
agree best with the calculation that considers an antiferromagnetic
ground state\cite{LDA3,LDA4}. On the other hand, this low energy
feature does not show up in dynamical mean field theory
calculation\cite{DMFT}.

\begin{figure*}[t]
\includegraphics[width=13cm]{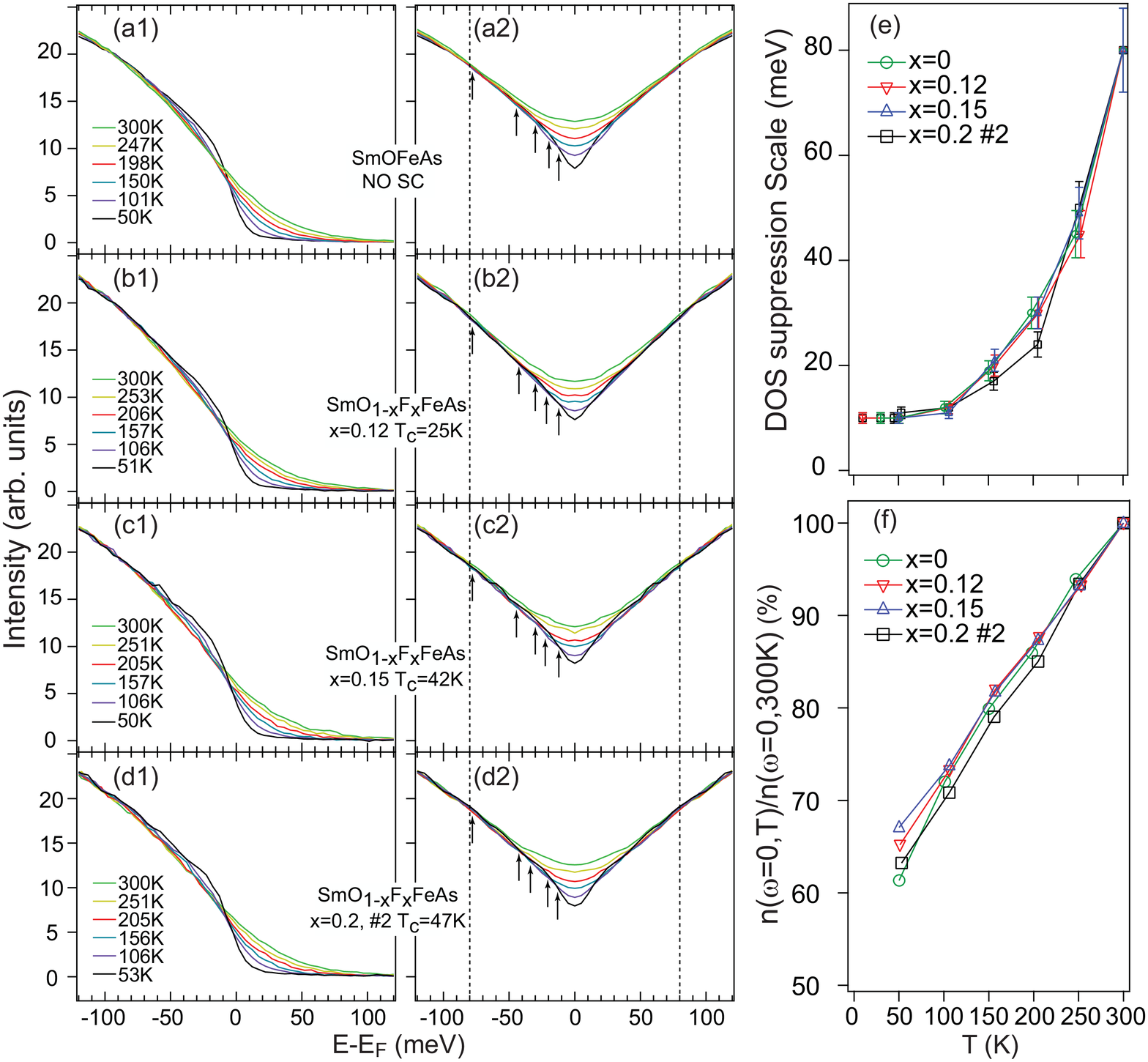}
\caption{(color online) Temperature dependence of the
SmO$_{1-x}$F$_{x}$FeAs AIPES spectrum near $E_F$ for (a1)$x=0$,
(b1)$x=0.12$, (c1) $x=0.15$, and (d1) $x=0.2$ respectively. The
spectra are symmetrized and shown in (a2-d2) respectively, and the
arrows indicate where further suppression of DOS occurs with
decreasing temperature. (e) Temperature dependence of the energy
scale of further spectral weight suppression [shown by arrows in
panels (a2-d2)]. The low temperature ones are taken from data in
Fig.4. (f) Temperature dependence of DOS at $E_F$ normalized by
their values at 300K. Data were taken with 22.7\,eV photons. }
\end{figure*}

The evolution of DOS near $E_F$ is studied as a function of doping
and temperature. As shown in Fig.3(a1-d1), the lineshape evolution
above and below $E_F$ clearly differs from the symmetrical
Fermi-Dirac distribution function normally observed on a
polycrystalline metal. The spectra [$n(\omega)$] are symmetrized to
remove the thermal broadening effects\cite{Norman}, and the
resulting $n(\omega)+n(-\omega)$ are shown in Fig.3(a2-d2)
respectively. This has been commonly practiced in the study of gap
in cuprate superconductors. In this way, a spectra weight
suppression with decreasing temperature is clearly revealed, which
occurs at the highest measured temperature over the energy range of
$\pm 80 meV$. With decreasing temperature, further suppressions of
the DOS happen in a smoothly shrinking energy window around $E_F$,
as shown by the arrows and summarized in Fig.3e. This resembles the
anisotropic pseudogap opening behavior in cuprates, where gap of
larger size opens at higher temperature\cite{Norman}. It also
indicates the maximum gap could well exceed 80\,meV. However, unlike
the cuprates, the suppression has no doping dependence. All AIPES
spectra exhibit similar behavior in the entire investigated doping
range. As quantified in Fig.3f, the DOS at $E_F$ has almost the same
linear temperature dependence when normalized by their values at
300K.

To study the superconducting state, high resolution AIPES spectra
were measured with 1\,meV steps near the Fermi energy at low
temperatures in Fig.\,4(a1-d1). The insets in Fig.4(a1-d1) show the
enlargement near $E_F$, the cross-points indicate a leading-edge gap
of $1\sim 4\,meV$ fluctuating with sample. The corresponding
symmetrized angle-integrated spectra are shown in Fig.\,4(a2-d2).
There is clearly an additional abrupt drop of DOS at a fixed
characteristic energy scale of 10\,meV at low temperatures (see
Fig.3e), which eventually causes a V-shaped DOS near $E_F$ at the
lowest temperatures. The 10 \,meV energy scale itself is quite
intriguing, since it is about $2 k_B T_c^{max}$ for
SmO$_{1-x}$F$_{x}$FeAs. Similar low energy scales of $2 k_B T_c$
have been observed for LOFFA\cite{takahashi,shin}, and
LOFFP\cite{shin}, therefore it is likely an intrinsic pseudogap
related to pairing fluctuation in the normal state. This gap might
have started to open above 100K, but the thermal broadening prevents
it being observed. If the pseudogap/superconducting gap transition
in optimally doped cuprates could apply here, it naturally evolves
into the superconducting gap once the system enters the
superconducting state. It is remarkable that such a pseudogap is
doping independent and very small in SOFFA, compared with the large
and doping dependent pseudogap in cuprates. Furthermore, the
V-shaped DOS with finite value at $E_F$ indicates an anisotropic gap
with nodes or Fermi arc in the superconducting state\cite{Ou1}.

\begin{figure}[t]
\includegraphics[width=8.5cm]{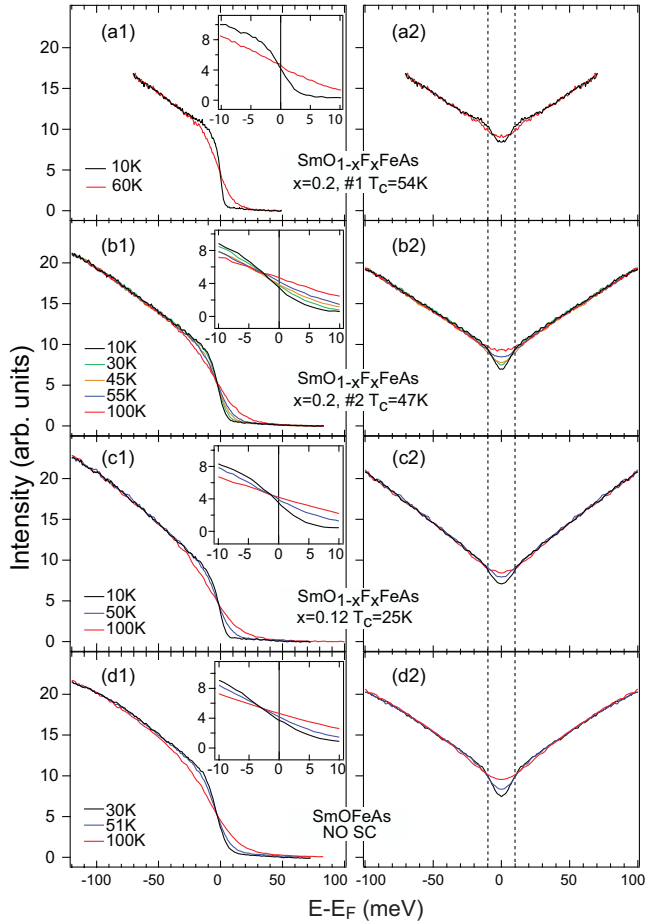}
\caption{(color online)  Temperature dependence of the
SmO$_{1-x}$F$_{x}$FeAs AIPES spectrum near $E_F$ at low temperatures
for (a1)$x=0.2$, $T_c=54K$, (b1) $x=0.2$, $T_c=47K$, (c1) $x=0.12$,
$T_c=25K$ and (d1) $x=0$ respectively. The corresponding symmetrized
spectra are shown in (a2-d2) respectively. Data in (b1-d1) were
taken with 22.7\,eV photons at SSRL, while data in (a1) were taken
with 8\,eV photons at HiSOR.}
\end{figure}

The doping-independent DOS suppressions do raise the question
whether they are artifacts from the polycrystalline nature of the
sample. Although the Meissner ratio is $50\sim60$\% for the
superconducting samples, as high as currently one could get, still
there are $40\sim50$\% non-superconducting part in the polycrystal,
which might come from the domain boundary or inhomogeneity of the
fluorine dopants or oxygen vacancies. It is not surprising if this
portion of the polycrystal could be independent of doping, and cause
the observed 30\% spectral weight suppression at $E_F$. Furthermore,
the quite large DOS around $E_F$ at 10K could be also attributed to
the non-superconducting portion besides the possible nodes or Fermi
arc of the superconducting portion, which naturally explains the
fluctuating leading edge gaps as well. However, for the suppression
at 10\,meV energy scale, its correlation with $T_c^{max}$ in several
systems indicates that it should be an intrinsic effects from the
bulk part of the polycrystal. For the suppression at 80\,meV energy
scale, there is no strong evidence to exclude the extrinsic
polycrystalline effects, except the fact that this energy scale is
smoothly connected to the 10\,meV scale as shown in Fig.3e.


To summarize, we have found doping independent behavior of the
spectral weight suppression in SmO$_{1-x}$F$_{x}$FeAs. A large
``pseudogap" of 80\,meV is observed from the highest measured
temperature, whose origin is currently unclear, and debatably, could
be extrinsic. A smaller gap of 10\,meV becomes observable below
100K, which is likely an intrinsic pseudogap in the normal state,
and would become a superconducting gap once in the superconducting
state. Therefore, the pseudogap behavior in iron oxypnictides is
very different from that in cuprate superconductors. Moreover, our
data alert that the results obtained from polycrystalline iron
oxypnictides have to be interpreted with caution.

The authors thank Dr. D. H. Lu and R. H. He for their kind help at
SSRL, and thank Prof. Z. D. Wang, and S. Y. Li for stimulating
discussions. This work was supported by the Nature Science
Foundation of China and by the Ministry of Science and Technology of
China (National Basic Research Program No.2006CB921300 and
2006CB922005), and STCSM of China. SSRL is operated by the DOE
Office of Basic Energy Science under Contract No. DE-AC03-765F00515.

Note added: During the preparation of this manuscript, we noticed
another work on SmO$_{1-x}$F$_{x}$FeAs is posted
online\cite{XJZhou}.

\end{document}